\documentclass[useAMS,usenatbib]{mn2e}

\usepackage[T1]{fontenc}
 \usepackage{pslatex}
\usepackage{amsmath}
\usepackage{amsfonts}
\usepackage{color}
\usepackage{url}
\usepackage{bm}
\usepackage{graphicx}
\usepackage{subfigure}
\usepackage{amssymb}
\usepackage{soul}
\usepackage{booktabs}
\usepackage{array}
\usepackage[utf8]{inputenc}
\inputencoding{latin1}
\inputencoding{utf8}

{\newif\ifnotend
\notendtrue
\def\veclist{ABCDEFGHIJKLMNOPQRSTUVWXYZabcdefghijklmnopqrstuvwxyz.}
\def\top#1#2.{#1}
\def\tail#1#2.{#2.}
\loop\expandafter\xdef\csname v\expandafter\top\veclist\endcsname%
{{\noexpand\bf\expandafter\top\veclist}}
\edef\veclist{\expandafter\tail\veclist}
\if\veclist.\notendfalse\fi\ifnotend\repeat}
\def\e{{\rm e}}

\def\pa{\partial}
\def\E{{\cal E}}

\mathchardef\mhyphen="2D

\title[Non-thermal spectra in relativistic sources]{Magnetic energy dissipation and origin of non-thermal spectra in radiatively efficient relativistic sources}
\author[Sobacchi \& Lyubarsky]{E. Sobacchi$^{1,2,3}$\thanks{E-mail: sobacchi@post.bgu.ac.il} and Y. E. Lyubarsky$^1$\thanks{E-mail: lyub@bgu.ac.il}\\
$^1$ Physics Department, Ben-Gurion University, P.O.B. 653, Beer-Sheva 84105, Israel \\
$^2$ Department of Natural Sciences, The Open University of Israel, P.O.B. 808, Raanana 4353701, Israel \\
$^3$ Department of Astronomy and Columbia Astrophysics Laboratory, Columbia University, New York, NY 10027, USA
}
\begin{document}

\date{}

\def\p{\partial}
\def\E{\textbf{E}}
\def\B{\textbf{B}}
\def\v{\textbf{v}}
\def\j{\textbf{j}}
\def\s{\textbf{s}}
\def\e{\textbf{e}}

\newcommand{\di}{\mathrm{d}}
\newcommand{\bfx}{\mathbf{x}}
\newcommand{\bfe}{\mathbf{e}}
\newcommand{\vlos}{\mathrm{v}_{\rm los}}
\newcommand{\Tspin}{T_{\rm s}}
\newcommand{\Tb}{T_{\rm b}}
\newcommand{\degree}{\ensuremath{^\circ}}
\newcommand{\Th}{T_{\rm h}}
\newcommand{\Tc}{T_{\rm c}}
\newcommand{\bfr}{\mathbf{r}}
\newcommand{\bfv}{\mathbf{v}}
\newcommand{\bfu}{\mathbf{u}}
\newcommand{\pc}{\,{\rm pc}}
\newcommand{\kpc}{\,{\rm kpc}}
\newcommand{\Myr}{\,{\rm Myr}}
\newcommand{\Gyr}{\,{\rm Gyr}}
\newcommand{\kms}{\,{\rm km\, s^{-1}}}
\newcommand{\de}[2]{\frac{\partial #1}{\partial {#2}}}
\newcommand{\cs}{c_{\rm s}}
\newcommand{\rb}{r_{\rm b}}
\newcommand{\rqu}{r_{\rm q}}
\newcommand{\bfOmega}{\pmb{\Omega}}
\newcommand{\bfOmegap}{\pmb{\Omega}_{\rm p}}
\newcommand{\bfXi}{\boldsymbol{\Xi}}

\maketitle

\begin{abstract}
The dissipation of turbulent magnetic fields is an appealing scenario to explain the origin of non-thermal particles in high-energy astrophysical sources. However, it has been suggested that the particle distribution may effectively thermalise when the radiative (synchrotron and/or Inverse Compton) losses are severe. Inspired by recent PIC simulations of relativistic turbulence, which show that electrons are impulsively heated in intermittent current sheets by a strong electric field aligned with the local magnetic field, we instead argue that in plasmas where the particle number density is dominated by the pairs (electron-positron and electron-positron-ion plasmas):
(i) as an effect of fast cooling and of different injection times, the electron energy distribution is ${\rm d}n_e/{\rm d}\gamma\propto\gamma^{-2}$ for $\gamma\lesssim\gamma_{\rm heat}$ (the Lorentz factor $\gamma_{\rm heat}$ being close to the equipartition value), while the distribution steepens at higher energies;
(ii) since the time scales for the turbulent fields to decay and for the photons to escape are of the same order, the magnetic and the radiation energy densities in the dissipation region are comparable;
(iii) if the mass energy of the plasma is dominated by the ion component, the pairs with a Lorentz factor smaller than a critical one (of the order of the proton-to-electron mass ratio) become isotropic, while the pitch angle remains small otherwise.
The outlined scenario is consistent with the typical conditions required to reproduce the Spectral Energy Distribution of blazars, and allows one to estimate the magnetisation of the emission site.
Finally, we show that turbulence within the Crab Nebula may power the observed gamma-ray flares if the pulsar wind is nearly charge-separated at high latitudes.
\end{abstract}



\section{Introduction}
\label{sec:introduction}

In the most extreme astrophysical environments (e.g. in the vicinity of black holes and pulsars), plasmas are expected to be in a regime where the electromagnetic energy largely exceeds the particles mass energy. Though it has long been clear from observations that these plasmas must emit non-thermal radiation copiously, the physical mechanism responsible for the dissipation of the electromagnetic energy and the consequent heating/acceleration of the non-thermal particles is far less understood.

Since shocks are inefficient in highly magnetised flows \citep[e.g.][]{KennelCoroniti1984}, turbulence is a natural candidate to dissipate the electromagnetic energy and consequently to heat/accelerate the particles. Turbulence in relativistic magnetised plasmas has long been studied from a fluid perspective \citep[e.g.][]{ThompsonBlaes1998, Cho2005, ZrakeMacfadyen2012, ZrakeMacfadyen2013, Zrake2014, ChoLazarian2014, ZrakeEast2016, TakamotoLazarian2016, TakamotoLazarian2017}. However, investigating the kinetic aspects of the problem has been hampered by the enormous required computational effort, and the actual mechanism that heats/accelerates the particles has therefore remained obscure.

Only recently, Particle-In-Cell (PIC) simulations are beginning to investigate the acceleration of particles in relativistic magnetised turbulence. \citet{ComissoSironi2018, ComissoSironi2019} studied the case of spontaneously decaying turbulence (namely, without any external driver continuously injecting energy into the system) in pair plasmas. In our opinion, some of the most remarkable results of these studies are:
\begin{enumerate}
\item Most of the electrons from the thermal background are impulsively heated up to a Lorentz factor $\gamma_{\rm heat}\sim B^2/8\pi n_e m_e c^2$ by a strong electric field aligned with the local magnetic field in intermittently formed, large-scale current sheets.
\item Electrons are subsequently accelerated into a non-thermal distribution extending up to $\gamma\gg\gamma_{\rm heat}$ by the stochastic interactions with the turbulent fluctuations. The electron energy spectrum observed in the simulations depends on the detailed properties of the plasma (e.g. magnetisation and amplitude of the magnetic field fluctuations compared to the background).
\item The typical pitch angle increases with the electron energy. The electrons with $\gamma\sim\gamma_{\rm heat}$ (primarily heated by the parallel electric field) have a small pitch angle, while the electrons with $\gamma\gg\gamma_{\rm heat}$ (primarily accelerated by the turbulent scattering) have a larger pitch angle.
\end{enumerate}
\citet{Zhdankin+2017, Zhdankin+2018b, Zhdankin+2018} and \citet{Wong+2019} studied the case of externally driven turbulence (namely, with a continuous energy injection into the system). They found the acceleration phase (item ii) to be similar to the studies of \citet{ComissoSironi2018, ComissoSironi2019}, while they did not find any clear evidence of an injection phase (item i). It is possible that the impact of the parallel electric field during the injection phase increases with the initial magnetisation of the plasma, which was indeed systematically higher in the studies of \citet{ComissoSironi2018, ComissoSironi2019} with respect to the studies of \citet{Zhdankin+2017, Zhdankin+2018b, Zhdankin+2018} (for a more extended discussion on the dependence of the work done by the parallel electric field on the magnetisation, see also Section 4.8 of \citealt{Zhdankin+2019}). Being interested in the case of strongly magnetised plasmas, one should therefore consider the impulsive electron heating during injection phase as discussed by \citet{ComissoSironi2018, ComissoSironi2019}.

The studies of \citet{ComissoSironi2018, ComissoSironi2019} and \citet{Zhdankin+2017, Zhdankin+2018b, Zhdankin+2018} have drawn attention to the role of turbulence in the heating/acceleration of non-thermal particles. However, it is not clear whether a population of non-thermal particles may survive in the common case when the radiative losses are severe. The effect of a fast (synchrotron and Inverse Compton) cooling has been included in an analytical model of turbulent plasmas by \citet{Uzdensky2018}. Assuming that the turbulent heating rate of individual electrons remains approximately constant over the entire dynamical time, and therefore neglecting the role of the impulsive electron heating during the injection phase, a steady state is reached where heating and cooling balance for any individual electron. As a consequence of this, a quasi-thermal electron energy distribution forms at a Lorentz factor $\gamma\sim 1/\sqrt{\tau_{\rm T}}$, where $\tau_{\rm T}\ll 1$ is the Thompson's optical depth of the system. Including the effect of Inverse Compton cooling in their PIC simulation of driven turbulence, \citet{Zhdankin+2019} have recently confirmed that a quasi-thermal electron distribution may be produced. However, observations of relativistic astrophysical plasmas usually show power-law photon spectra, indicating that the electron distribution has a similar scaling.

Here we show that the impulsive electron heating in intermittent, large-scale current sheets may be a key element to produce a non-thermal electron population even when the radiative losses are severe. Motivated by the results of the PIC simulations of \citet{ComissoSironi2018, ComissoSironi2019}, we discuss the scenario where most of the pairs are impulsively heated up to equipartition ($\gamma\sim\gamma_{\rm heat}$) while entering a current sheet, after which they cool down for the rest of the dynamical time. As an effect of cooling and of different injection times, the electron energy distribution at $\gamma\lesssim\gamma_{\rm heat}$ is a power-law of the form ${\rm d}n_e/{\rm d}\gamma \propto \gamma^{-2}$, which is consistent with blazar observations. Since most of the work during the injection phase is done by a parallel electric field, the momentum distribution is (at least initially) strongly elongated in the direction of the magnetic field.

This paper is organised as follows. In Section \ref{sec:heating}, we discuss the injection phase. In Section \ref{sec:spectrum}, we show how radiative cooling shapes the electron energy distribution. In Section \ref{sec:astro}, we discuss the astrophysical applications of the outlined scenario, focusing on the modelling of the non-thermal emission from galactic jets and of the gamma-ray flares from the Crab Nebula. Finally, in Section \ref{sec:conclusions} we summarise our conclusions.

\section{Electron heating in turbulent plasmas}
\label{sec:heating}

We consider a plasma where the particle number density is dominated by the pairs (electron-positron or electron-positron-ion plasma), characterised by a pair number density $n_e\equiv n_{e^-}+n_{e^+}$ and a proton number density $n_p\equiv n_{e^-}-n_{e^+} \ll n_e$. As explained in more detail in Section \ref{sec:astro}, astrophysical plasmas are expected to be significantly loaded by pairs in different environments, including galactic jets and Pulsar Wind Nebulae.
We assume that the plasma is threaded by a magnetic field $B$.  We are interested in the case when the turbulence is strong, in the sense that the amplitude of the turbulent fluctuations is $\delta B\sim B$. It is well known that the turbulent component of the magnetic field is dissipated on a typical time scale $t_{\rm dyn}\sim L/v_{\rm A}$, where $L$ is the scale of the largest turbulent eddy and $v_{\rm A}$ is the Alfv\'{e}n velocity \citep[e.g.][]{Biskamp2003}. In the following we assume that $L$ is comparable to the size of the system, and that $v_{\rm A}\sim c$ as appropriate in the relativistic limit.

We assume the kinetic energy of the particles prior to dissipation to be at most mildly relativistic ($\gamma_{\rm in}\beta_{\rm in}\lesssim 1$). We consider the case when the initial magnetisation of the pairs,
\begin{equation}
\sigma_{0e} \sim \frac{B^2}{8\pi n_e m_e c^2} \;,
\end{equation}
is much larger than unity ($\sigma_{0e}\gg 1$). If the ions are present, the initial magnetisation of the plasma is
\begin{equation}
\sigma_0\sim \frac{B^2}{8\pi (n_e m_e +n_p m_p)c^2} \sim \frac{\sigma_{0e}}{1+n_pm_p/n_em_e} \;.
\end{equation}
Note that $\sigma_0$ may be significantly smaller than $\sigma_{0e}$ if the ions dominate the mass density of the system, namely $n_pm_p\gtrsim n_em_e$. In this case, we assume that $\sigma_0\gtrsim 1$, always keeping $\sigma_{0e}\gg 1$.

\subsection{Lorentz factor of the heated electrons}
\label{sec:gamma_heat}


After the turbulent component of the magnetic field has been dissipated, electrons and positrons are heated up to a typical Lorentz factor $\gamma_{\rm heat}\gg\gamma_{\rm in}$, which can be estimated from the principle of energy conservation. The energy density of the heated pairs, $U_e\sim\gamma_{\rm heat}n_em_ec^2$, is equal to a fraction $\varepsilon_e<1$ of the available magnetic energy density, $U_B\sim B^2/8\pi$ (the remaining fraction $\varepsilon_p=1-\varepsilon_e$ heats the protons). This gives
\begin{equation}
\label{eq:gamma_heat}
\gamma_{\rm heat}\sim \varepsilon_e\frac{B^2}{8\pi n_e m_e c^2} \sim \varepsilon_e \sigma_{0e}\;.
\end{equation}
In the case of a pair plasma, one has $\gamma_{\rm heat}\sim\sigma_{0e}$ since there are no protons and therefore $\varepsilon_e=1$. In the following we argue that the efficiency $\varepsilon_e$ is of order unity, and therefore $\gamma_{\rm heat}\sim\sigma_{0e}$, also in electron-positron-ion plasmas.

As discussed in the Introduction, recent PIC simulations suggest that electrons are heated from their initial $\gamma_{\rm in}$ up to $\gamma_{\rm heat}$ in the course of one single event, while passing through a large-scale current sheet where the anti-parallel component $\delta B\sim B$ of the magnetic field reconnects \citep[][]{ComissoSironi2018, ComissoSironi2019}. Recently, \citet{Petropoulou+2019} have studied reconnection in electron-positron-ion plasmas, showing that the post-reconnection energy is shared roughly equally between magnetic fields, pairs, and protons (more specifically, they found that $\varepsilon_e\sim 1/3$ if $\sigma_{0e}\gtrsim 3$). Hence, we expect the efficiency $\varepsilon_e$ to be of order unity in the regime $\sigma_{0e}\gg 1$ that we are interested in.

\subsection{Heating rate}

Since the turbulent component of the magnetic field is dissipated over a dynamical time, $t_{\rm dyn}\sim L/c$, the heating rate per electron, $P_{\rm heat}$, needs to satisfy the following equation:
\begin{equation}
\label{eq:heating}
\gamma_{\rm heat} m_e c^2 \sim \int_0^{t_{\rm dyn}} P_{\rm heat} {\rm d}t \;.
\end{equation}
As discussed in the Introduction (for more details, see also Appendix \ref{sec:appA}), pairs are heated by an electric field $E_\parallel$ aligned with the local magnetic field, arising as the plasma becomes starved of free charges in intermittently formed current sheets. If the velocity of the plasma flowing into the reconnection layer is at least mildly relativistic, one expects the electric field to be of the same order of magnitude of the magnetic field, namely $E_\parallel\sim B$ (in their PIC simulations, \citet{ComissoSironi2018, ComissoSironi2019} found a typical $E_\parallel\sim 0.1\times B$). Pairs are therefore heated on a time scale 
\begin{equation}
t_{\rm heat}\sim \frac{\gamma_{\rm heat}m_e c}{eB}\sim \frac{\gamma_{\rm heat}}{\gamma_{\rm max}}t_{\rm dyn}\;,
\end{equation}
where $\gamma_{\rm max}m_ec^2\sim eBL$ is the maximum energy allowed by the size of the system. Note that if $\gamma_{\rm heat}\ll\gamma_{\rm max}$ the turbulent heating becomes approximately impulsive (namely, $t_{\rm heat}\ll t_{\rm dyn}$). We may therefore estimate the turbulent heating rate per electron as
\begin{equation}
\label{eq:Pheat}
P_{\rm heat}\sim \frac{\gamma_{\rm heat}m_e c^2}{t_{\rm heat}} \sim \frac{\gamma_{\rm max}m_e c^2}{t_{\rm dyn}}
\end{equation}
if $\left|t-t_{\rm inj}\right|\lesssim t_{\rm heat}/2$ and $P_{\rm heat}\sim 0$ otherwise. In our model different electrons have different injection times, with $t_{\rm inj}$ being uniformly distributed over the range $0\lesssim t_{\rm inj}\lesssim t_{\rm dyn}$.

In the model of \citet[][]{Uzdensky2018} the heating is instead slow, in the sense that the heating rate per electron remains approximately constant over the entire dynamical time. Hence, from Eq. \eqref{eq:heating} one finds that $P_{\rm heat}\sim\gamma_{\rm heat}m_e c^2 / t_{\rm dyn}$.

\section{Cooling and electron energy distribution}
\label{sec:spectrum}

\subsection{Radiative cooling prescription}

Besides being heated due to the turbulent energy dissipation, electrons cool down due to synchrotron and Inverse Compton emission.
We parametrise the radiated power per electron as
\begin{equation}
\label{eq:Pcool}
P_{\rm cool}\sim \frac{\gamma^2 m_e c^2}{\gamma_{\rm cool}t_{\rm dyn}} \;,
\end{equation}
where $\gamma_{\rm cool}$ is defined as the Lorentz factor of the electrons that cool approximately within a dynamical time. The parametrisation of Eq. \eqref{eq:Pcool} is motivated by the well-known scaling $P_{\rm cool}\propto \gamma^2$ \citep[e.g.][]{RybickiLightman1979}, and by the fact that the cooling time, $t_{\rm cool}\sim \gamma m_e c^2/P_{\rm cool}\sim t_{\rm dyn}\gamma_{\rm cool}/\gamma$, is indeed equal to the dynamical time, $t_{\rm dyn}$, when $\gamma\sim\gamma_{\rm cool}$.

\subsection{Electron energy distribution}
\label{sec:regimes}

\begin{figure}
\centering
\includegraphics[width=0.47\textwidth]{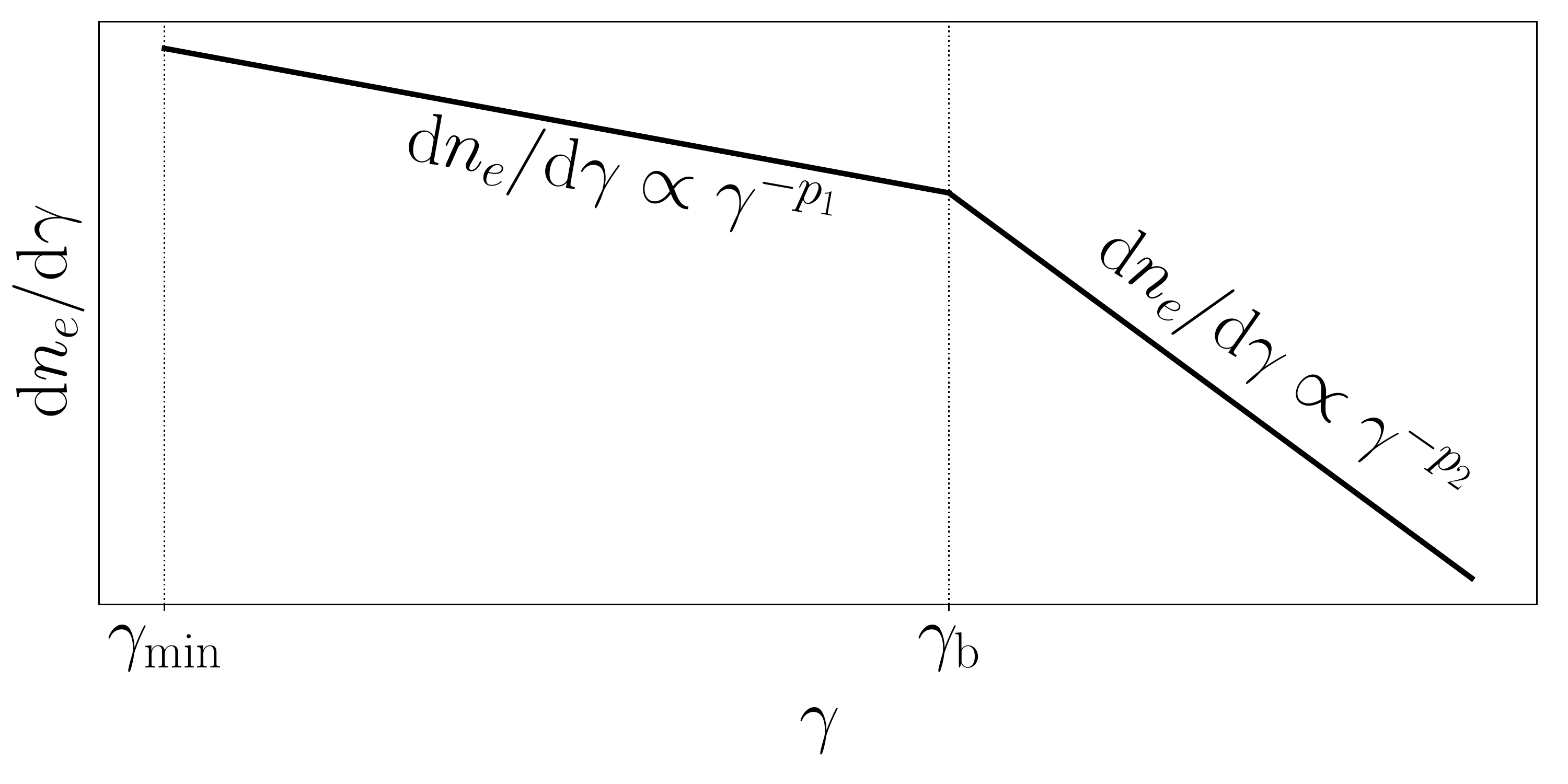}
\includegraphics[width=0.47\textwidth]{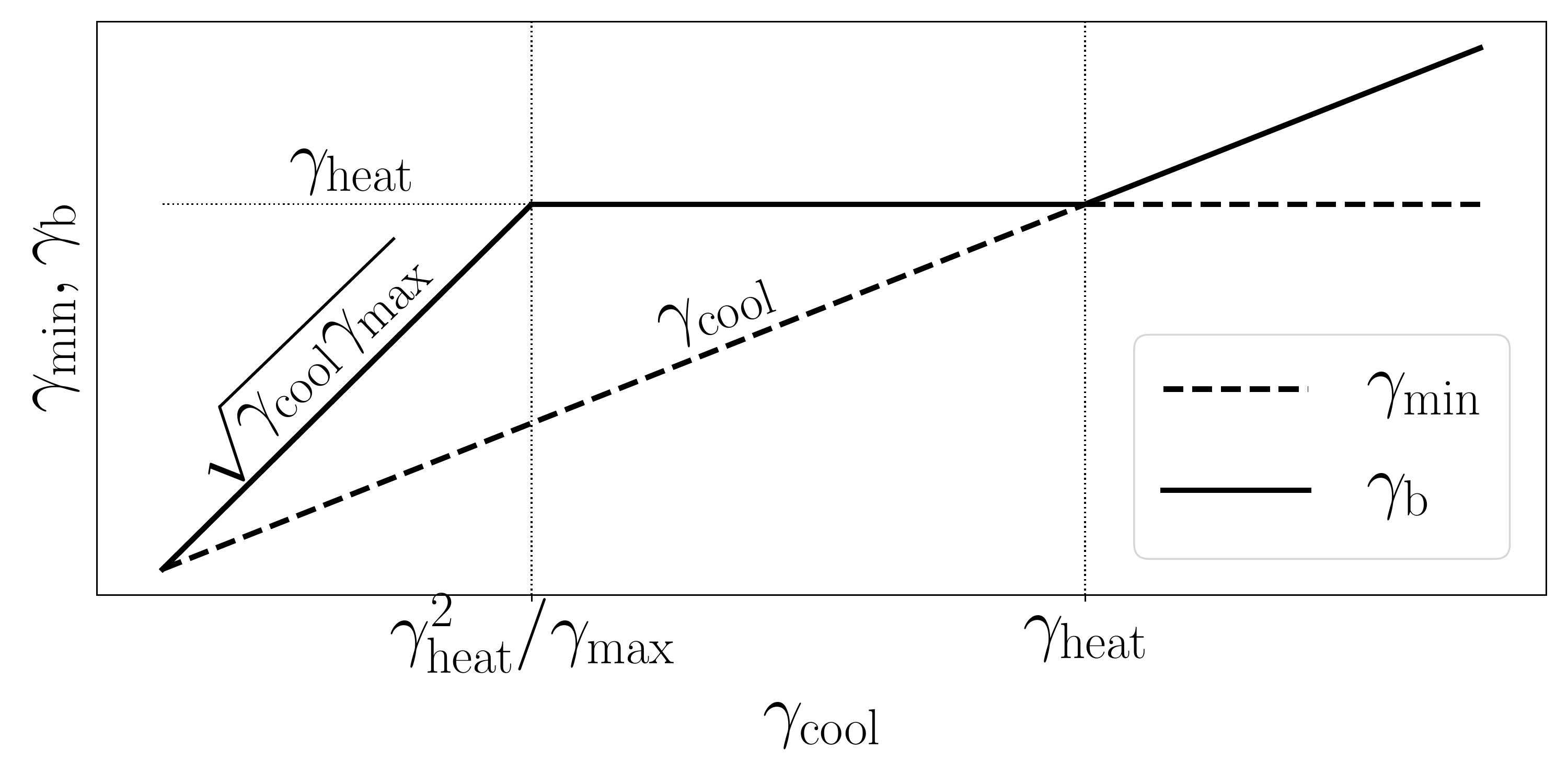}
\caption{Top panel: illustrative sketch of the electron energy distribution produced by the turbulent dissipation of the magnetic energy, including the effect of radiative cooling. Bottom panel: minimum and break Lorentz factors of the distribution. The Lorentz factor $\gamma_{\rm heat}$ has the same order of magnitude of the initial magnetisation of the electrons, namely $\gamma_{\rm heat}\sim\sigma_{0e}\sim B^2/8\pi n_em_ec^2\gg 1$; $\gamma_{\rm cool}$ is the Lorentz factor of the electrons radiating most of their energy within one dynamical time; $\gamma_{\rm max}$ is the maximum electron Lorentz factor allowed by the size of the system. In the fast cooling regime ($\gamma_{\rm cool}\lesssim\gamma_{\rm heat}$) the power-law index below the break is $p_1=2$.}
\label{fig:spectrum}
\end{figure}

The electron energy distribution may be approximated as a broken power-law extending over $\gamma\gtrsim\gamma_{\rm min}$. Both the minimum Lorentz factor, $\gamma_{\rm min}$, and the break Lorentz factor, $\gamma_{\rm b}$, are determined by the three fundamental energy scales $\gamma_{\rm heat}$, $\gamma_{\rm cool}$, and $\gamma_{\rm max}$. The main results are discussed in the following and summarised in Figure \ref{fig:spectrum}.

\subsubsection{Ultra fast cooling regime ($\gamma_{\rm cool}\lesssim\gamma_{\rm heat}^2/\gamma_{\rm max}$)}

In this regime the radiative losses in the reconnection layer are important. Indeed, cooling and heating in the reconnection layer balance each other when $\gamma\sim\sqrt{\gamma_{\rm cool}\gamma_{\rm max}}$ (this is obtained from the condition $P_{\rm heat}\sim P_{\rm cool}$ using Eqs. \ref{eq:Pheat} and \ref{eq:Pcool}). If $\gamma_{\rm heat}\gtrsim\sqrt{\gamma_{\rm cool}\gamma_{\rm max}}$, which is equivalent to $\gamma_{\rm cool}\lesssim\gamma_{\rm heat}^2/\gamma_{\rm max}$, the bulk of the electrons can only be heated up to $\gamma_{\rm b}\sim\sqrt{\gamma_{\rm cool}\gamma_{\rm max}}$.

Electrons start cooling down after leaving the reconnection layer, and their Lorentz factor reaches $\gamma_{\rm min}\sim\gamma_{\rm cool}$ within a dynamical time. Due to the difference in the injection times, a non-thermal energy spectrum ${\rm d}n_e/{\rm d}\gamma\propto\gamma^{-2}$ is produced over the range of Lorentz factors $\gamma_{\rm min}\lesssim\gamma\lesssim\gamma_{\rm b}$. The spectrum produced by radiative cooling is a power-law of the form ${\rm d}n_e/{\rm d}\gamma\propto\gamma^{-2}$ because the number of electrons per unit energy, $\gamma\times {\rm d}n_e/{\rm d}\gamma$, is proportional to the cooling time at that energy, $t_{\rm cool}\propto\gamma^{-1}$.

Since the electric field inside the reconnection layer cannot heat the electrons beyond $\gamma\sim\gamma_{\rm b}$, one generically expects the spectrum to steepen at $\gamma\gtrsim\gamma_{\rm b}$. The shape of the spectrum at $\gamma\gtrsim\gamma_{\rm b}$ likely depends on the detailed properties of the plasma, and its extension may be severely limited by the effect of cooling (see the discussion in Appendix \ref{sec:appB}). Finally, note that the condition $\gamma_{\rm cool}\lesssim\gamma_{\rm heat}^2/\gamma_{\rm max}$ becomes very restrictive if $\gamma_{\rm max}$ is large.

\subsubsection{Moderately fast cooling regime ($\gamma_{\rm heat}^2/\gamma_{\rm max}\lesssim\gamma_{\rm cool}\lesssim\gamma_{\rm heat}$)}

In this regime the radiative losses within the reconnection layer can be neglected, while the remaining physics is similar to the ultra fast cooling regime discussed above. One finds a energy spectrum of the form ${\rm d}n_e/{\rm d}\gamma\propto\gamma^{-2}$ extending over $\gamma_{\rm min}\lesssim \gamma \lesssim \gamma_{\rm b}$ and steepening at $\gamma \gtrsim \gamma_{\rm b}$, with $\gamma_{\rm min}\sim\gamma_{\rm cool}$ and $\gamma_{\rm b}\sim\gamma_{\rm heat}$. As in the ultra fast cooling regime, the extension of the distribution at $\gamma\gtrsim\gamma_{\rm b}$ may be severely limited by the effect of cooling.

It is important to realise that in the fast cooling regime the pairs may retain a significant magnetisation after the turbulent component of the magnetic field has been dissipated. Indeed the kinetic energy density of the pairs is $U_e\sim \int_{\gamma_{\rm cool}}^{\gamma_{\rm heat}} {\rm d}\gamma \; \gamma m_e c^2 \; ({\rm d}n_e/{\rm d}\gamma) \sim \gamma_{\rm cool}n_e m_e c^2 \log(\gamma_{\rm heat}/\gamma_{\rm cool})\lesssim \gamma_{\rm heat}n_e m_e c^2\sim U_B$. Most of the dissipated magnetic energy is therefore converted into radiation within a dynamical time. Since the escape time of the photons from the dissipation region is also $\sim t_{\rm dyn}$, one expects the magnetic and the radiation energy density to be comparable ($U_B\sim U_\gamma$) if the external sources of photons can be neglected.

\subsubsection{Slow cooling regime ($\gamma_{\rm heat}\lesssim\gamma_{\rm cool}$)}

In this regime the electrons that are heated up to $\gamma\sim\gamma_{\rm heat}$ do not cool within a dynamical time. Hence, one finds that $\gamma_{\rm min}\sim\gamma_{\rm heat}$. If turbulence accelerates the electrons in a non-thermal tail extending up to $\gamma\gg\gamma_{\rm heat}$, one expects a spectral break break to appear at $\gamma_{\rm b}\sim\gamma_{\rm cool}$ as an effect of cooling. In this regime, the energy spectrum likely depends on the detailed properties of the plasma.

\subsection{Anisotropy of the distribution}
\label{sec:anisotropy}

In the fast cooling regime, most the electrons are impulsively heated by a parallel electric field before cooling down for the rest of the dynamical time. Since the IC scattering does not change the pitch angle significantly while the electron cools (see Appendix B of \citealt{TavecchioSobacchi2019}), the momentum of all the electrons with $\gamma\lesssim\gamma_{\rm heat}$ may remain aligned with the magnetic field in the proper frame of the plasma.

A possible exception, which is relevant only for electron-positron-ion plasmas, is the following. \citet{SobacchiLyubarsky2019} have recently proposed that the pitch angle is gyro-resonantly scattered by Alfv\'en waves (not belonging to the turbulent cascade) that grow unstable if the jet has a proton component that dominates the mass density ($n_pm_p\gtrsim n_em_e$).\footnote{The electrons pitch angle may also be gyro-resonantly scattered by the turbulent fluctuations. These fluctuations are characterised by a scale-dependent anisotropy, namely $l_\perp/l_\parallel\sim (l_\parallel/L)^\alpha$ where $1/2\lesssim\alpha\lesssim 1$ (the exact value of $\alpha$ depends on the model), and $l_\perp$ ($l_\parallel$) is the the scale of the fluctuation in the direction perpendicular (parallel) to the magnetic field \citep[e.g.][]{GoldreichSridhar1995, ThompsonBlaes1998, Boldyrev2006}. Gyro-resonant scattering by the turbulent fluctuations requires that (i) the electron travel a distance comparable to $l_\parallel$ in a Larmor time, which gives the usual resonance condition $\gamma m_e c^2\sim eBl_\parallel$; (ii) the electron Larmor radius is smaller than $l_\perp$, which gives $\gamma m_e c^2\theta\lesssim eBl_\perp$ where $\theta$ is the pitch angle in the proper frame of the plasma \citep[e.g.][]{Chandran2000}. Putting all together, one finds that $\theta\lesssim l_\perp/l_\parallel\sim  (l_\parallel/L)^\alpha \sim (\gamma/\gamma_{\rm max})^\alpha$, namely $\theta$ increases with the electron energy. If $\gamma_{\rm heat}\ll\gamma_{\rm max}$ and gyro-resonant scattering by the turbulent fluctuations is the only relevant process, the pitch angle of the bulk of the electrons remains small.} This process works only for electrons with a Lorentz factor smaller than a critical $\gamma_{\rm iso}$, which is of the order of the proton-to-electron mass ratio $m_p/m_e$.
Hence, we suggest that (i) if the jet has a significant proton component ($n_pm_p\gtrsim n_em_e$) and the initial magnetisation of the pairs is smaller than $\gamma_{\rm iso}$ ($\sigma_{0e}\lesssim \gamma_{\rm iso}$), the distribution becomes approximately isotropic; (ii) if the jet has a significant proton component ($n_pm_p\gtrsim n_em_e$) and the initial magnetisation of the pairs is $\sigma_{0e}\gtrsim \gamma_{\rm iso}$, the electrons with $\gamma\lesssim \gamma_{\rm iso}$ are isotropised, while the electrons with $\gamma_{\rm iso}\lesssim\gamma\lesssim\gamma_{\rm heat}$ remain anisotropic; (iii) if the jet composition is dominated by pairs ($n_p m_p\lesssim n_e m_e$), the distribution remains anisotropic for all $\gamma\lesssim\gamma_{\rm heat}$.

Finally, note that if the protons are heated significantly (see the discussion in the last paragraph of Section \ref{sec:gamma_heat}), after the turbulence has decayed the total magnetisation of the plasma is of order unity ($\sigma\sim 1$) and the bulk motions of the plasma are therefore only mildly relativistic. Hence, the anisotropy of the distribution is approximately the same both in the frame of the dissipation region and in the proper frame of the plasma.

\section{Astrophysical implications}
\label{sec:astro}

\subsection{Blazar jets}

The super massive black holes residing in the centre of galaxies are able to launch jets that reach relativistic velocities. Blazars are thought to be galactic jets pointing in the direction of the Earth \citep[e.g.][]{UrryPadovani1995}. Since the radiation from the jet is strongly beamed (and often completely outshines the other components, such as the accretion disc), blazars are ideal natural laboratories to study the extreme energy dissipation regimes occurring in relativistic jets.
According to a widely accepted paradigm, galactic jets are launched hydromagnetically \citep[e.g.][]{Blandford1976, Lovelace1976, BlandfordZnajek1977}, with the jet's energy budget being initially dominated by the Poynting flux. Turbulent dissipation of the magnetic energy is therefore a natural candidate to explain the blazar emission.

The Spectral Energy Distribution (SED) of blazars is characterised by two broad non-thermal components, the first one peaking at IR-optical-UV frequencies, and the second one peaking in the $\gamma$ rays. The SED follows a well known sequence \citep[e.g.][]{Fossati1998, Ghisellini+2017}: the SED of the faintest objects peaks at higher frequencies, and the two components have comparable luminosities;  these objects usually show weak emission lines (if any), and are therefore classified as BL Lacs. The SED of the brightest objects peaks at lower frequencies, and the high energy component is more luminous; these objects often show strong emission lines, and are therefore classified as Flat Spectrum Radio Quasars (FSRQ). The first component of the SED is due to the synchrotron radiation from a population of non-thermal electrons, while in the context of leptonic models the second component is usually attributed to the Comptonization of either the synchrotron photons themselves (in the case of BL Lacs), or of an external photon field (in the case of FSRQ) \citep[e.g.][]{Sikora1994, Sikora1997, Sikora2009, Ghisellini1998, Ghisellini2010, Tavecchio1998}.

Reproducing the broad SED of blazars requires the non-thermal electron distribution to extend over several orders of magnitude in energy. This might be a serious problem if turbulence heats the electrons slowly \citep[as in the model of][]{Uzdensky2018}, since in this case a quasi-thermal electron distribution is expected when the cooling is fast. If instead the electrons are heated impulsively, a power-law distribution of the form ${\rm d}n_e/{\rm d}\gamma\propto\gamma^{-2}$ is expected over a broad range of energies ($\gamma_{\rm cool}\lesssim\gamma\lesssim\gamma_{\rm heat}$) as an effect of fast cooling (see the discussion in Section \ref{sec:regimes}), which is in better agreement with observations.

In the following we discuss the possible application of our scenario to the interpretation of the SED of FSRQ and BL Lacs. For a typical $B\sim 1{\rm\; G}$ and $L\sim 10^{15}{\rm\; cm}$ in the energy dissipation region, one finds that $\gamma_{\rm max}\sim 10^{12}$. Since in blazars typically $\gamma_{\rm heat}\lesssim 10^6$ (see below), the system has a huge dynamic range ($\gamma_{\rm heat}\ll\gamma_{\rm max}$). Since $\gamma_{\rm heat}^2/\gamma_{\rm max}\lesssim 1$, the system is in the moderately fast or slow cooling regime (see the discussion in Section \ref{sec:regimes}).

\subsubsection{Flat Spectrum Radio Quasars}

Modelling the SED of FSRQ indicates that (i) the dissipation region is magnetised, with the inferred magnetic and electron energy density being typically not far from equipartition, namely $U_B \sim U_e$; (ii) in most of the objects the bulk of the electrons cool efficiently, and are consistently distributed as ${\rm d}n_e/{\rm d}\gamma\propto\gamma^{-2}$ at the lowest energies \citep[e.g.][]{CelottiGhisellini2008, Ghisellini2010}. Such conditions may naturally arise in a relativistic turbulent magnetised plasma in the fast cooling regime (see Section \ref{sec:regimes}).

The magnetisation of the plasma can be estimated from the observed SED. The break Lorentz factor of the electron distribution is $\gamma_{\rm b}\sim 10^2$ when $\gamma_{\rm cool}\lesssim 10^2$, and $\gamma_{\rm b}\sim\gamma_{\rm cool}$ when $\gamma_{\rm cool}\gtrsim 10^2$ \citep[see Figure 3 of][]{Ghisellini2010}. This is consistent with our model (see the bottom panel of our Figure \ref{fig:spectrum}), and indicates that $\gamma_{\rm heat}\sim 10^2$. Since $\gamma_{\rm heat}$ can be used as a proxy for the magnetisation of the pairs in the dissipation region, we expect that $\sigma_{0e}\sim 10^2$.

Few attempts have been made to quantify the elusive proton component of the jet. \citet[][]{SikoraMadejski2000} suggested that $n_e/n_p\sim 10\mhyphen 100$ in order to produce the observed amount of soft X-ray radiation. \citet[][]{GhiselliniTavecchio2010} argued that $n_e/n_p\lesssim 10$, since in the opposite case the jet would decelerate too much while Compton scattering the external photons. More recently, different groups argued that matching the jet power inferred from the SED fitting and from radio lobe calorimetry requires that $n_e/n_p\sim 10\mhyphen 20$ \citep[e.g.][]{Kang+2014, Sikora2016, Pjanka+2017, Fan+2018}. These results suggest that in the dissipation region the magnetisation of the plasma (including the protons) is of order unity, namely $\sigma_0\sim (n_e m_e/n_p m_p)\sigma_{0e}\sim 1$ for a typical $\sigma_{0e}\sim 10^2$ and $n_e/n_p\sim 20$. Finally, since $n_p m_p\gtrsim n_e m_e$ and $\sigma_{0e}\lesssim m_p/m_e$, assuming an isotropic electron distribution may be acceptable (see Section \ref{sec:anisotropy}).

\subsubsection{BL Lacs}

In the case of BL Lacs, reproducing the SED also requires a power-law electron distribution of the form ${\rm d}n_e/{\rm d}\gamma \propto \gamma^{-2}$ extending over $\gamma\lesssim\gamma_{\rm b}$ \citep[e.g.][]{Tavecchio2010}. Moreover, since the synchrotron and the IC peaks of the SED have comparable luminosities, one would expect the magnetic and the radiation energy density in the dissipation region to be comparable, namely $U_B\sim U_\gamma$.

Even though both these conditions are naturally expected in a fast cooling turbulent plasma (see Section \ref{sec:regimes}), providing a fully convincing interpretation remains difficult. Indeed, two main inconsistencies arise: under the assumption that the electron distribution is isotropic, fitting the SED indicates that (i) $U_B\ll U_e$, so that the jet should be matter dominated even if the proton component is completely absent; (ii) $\gamma_{\rm cool}\gg\gamma_{\rm b}$, so that the role of cooling in shaping the electron energy distribution should be negligible (\citealt{TavecchioGhisellini2016}; see also \citealt{InoueTanaka2016, NalewajkoGupta2017, Costamante+2018}).\footnote{Note that, if the non-thermal electrons in BL Lacs are not cooling efficiently ($\gamma_{\rm cool}\gg\gamma_{\rm b}$) and the break is therefore not due to cooling, one would need an acceleration mechanism that produces (i) an electron energy distribution ${\rm d}n_e/{\rm d}\gamma \propto \gamma^{-2}$ at low energies; (ii) an energy distribution ${\rm d}n_e/{\rm d}\gamma \propto \gamma^{-p}$ with $p>2$ at high energies. To the best of our knowledge, no such acceleration mechanism has been proposed so far.}

As shown by \citet{SobacchiLyubarsky2019} and \citet{TavecchioSobacchi2019}, these inconsistencies may be solved if the momentum of the highest energy electrons remains nearly aligned with the magnetic field, in which case the SED may be reproduced under equipartition and fast cooling conditions (namely, $U_B\sim U_e$ and $\gamma_{\rm cool}\ll\gamma_{\rm b}$). Such a scenario appears possible for BL Lacs, where the typical break Lorentz factor of the electron distribution, $\gamma_{\rm b}\sim 10^3 \mhyphen 10^6$ \citep[e.g.][]{Tavecchio2010}, suggests a high magnetisation of the pairs, $\sigma_{0e}\sim 10^3 \mhyphen 10^6$. As discussed in Section \ref{sec:anisotropy}, in this case the effect of the anisotropy may be important since $\sigma_{0e}\gtrsim m_p/m_e$.

Though the presence of ions in BL Lac jets is supported by the recent detection of a high-energy neutrino associated with the BL Lac object TXS 0506+056 \citep[][]{TXS18}, quantifying the ion number density remains difficult. \citet[][]{Madejski+2016} argued that $n_e/n_p\gtrsim 30$ in the BL Lac object PKS 2155-304 in order to avoid an unreasonably large jet power. Taking the same fiducial value of $n_p/n_e$ for all BL Lacs, and using the fact that $\sigma_{0e}\gtrsim 10^3$, we estimate that $\sigma_0\sim (n_e m_e/n_p m_p)\sigma_{0e}\gtrsim 15$. This suggests a higher magnetisation in the energy dissipation region with respect to FSRQ.

\citet[][]{Nalewajko2016} considered relativistic magnetic reconnection as the energy dissipation mechanism acting in blazars. He suggested that the blazar sequence might be characterised by a systematic trend in the magnetisation of the jet (with lower $\sigma_0$ in FSRQ and higher $\sigma_0$ in BL Lacs) regulated by the efficiency of pair production. Our conclusions are consistent with this suggestion.

\subsection{Crab gamma-ray flares}

The rapid variability in the gamma-rays is now a well established property of the Crab Nebula \citep[e.g.][]{Tavani+2011, Abdo+2011, Buehler+2012}. The most spectacular April 2011 flare \citep[][]{Buehler+2012, Striani+2013, Weisskopf+2013} showed several puzzling features:
\begin{enumerate}
\item The flare lasted for $t_{\rm flare}\sim 9{\rm\; days}$. This corresponds to a typical scale $ct_{\rm flare}\sim 2\times 10^{16}{\rm\; cm}$, at least one order of magnitude smaller than the equatorial size of the wind termination shock, which is $\sim 0.1{\rm\; pc}$. Moreover, variability was observed on a time scale of $\sim 8{\rm\; hr}$ during the flare.
\item The emitted isotropic power above $100{\rm\; MeV}$ at the peak of the flare was $\mathcal{L}\sim 4\times 10^{36}{\rm\; erg\; s}^{-1}$, which is $\sim 1\%$ of the total spin-down power of the pulsar.
\item The spectrum of the flaring component appeared at the high-energy end of the persistent synchrotron emission, which cuts off near the synchrotron burn-off limit, $\varepsilon_{\rm bo}\sim 160{\rm\; MeV}$. The spectral peak was observed up to photon energies of $\varepsilon_{\rm p}\sim 400{\rm\; MeV}$, which is significantly beyond the classical $\varepsilon_{\rm bo}$.
\item The spectrum was unusually hard, requiring a power-law energy distribution ${\rm d}n_e/{\rm d}\gamma\propto \gamma^{-p}$ with $p\sim 1.5$, or even a mono-energetic distribution. This is inconsistent with the the steeper power laws ($p\gtrsim 2$) expected in shock acceleration \citep[e.g.][]{BlandfordEichler1987, Spitkovsky2008, SironiSpitkovsky2009}.
\item The cutoff energy of the flaring component and the luminosity above $100{\rm\; MeV}$ were correlated, the observed correlation being described by $\mathcal{L}\propto \varepsilon_{\rm p}^\alpha$ with $\alpha=3.42\pm 0.86$.
\end{enumerate}

In order to radiate near the burn-off limit, electrons need to be accelerated by an electric field comparable to the magnetic field. This suggests that the flares are produced in the polar region, where the post-shock flow remains strongly magnetised and the bulk motions of the plasma are therefore relativistic (e.g. \citealt{Lyubarsky2012, Komissarov2013}; see however \citealt{Bykov+2012, Zrake2016}). The post-shock flow initially expands and decelerates but eventually converges because the magnetic hoop stress is not counterbalanced by either the poloidal field or the plasma pressure. Turbulence may develop as a consequence of the instability of such converging flows \citep[e.g.][]{SobacchiLyubarsky2018}.

In our opinion, it is crucial to understand whether a turbulent plasma may radiate most of its energy near the burn-off limit $\varepsilon_{\rm bo}$. If this is the case, the appearance of flares peaking a factor of a few above $\varepsilon_{\rm bo}$ may be due to the intermittent beaming of the accelerated electrons, which is an expected property of turbulent systems \citep[e.g.][]{Zhdankin+2019}. Variations in the relativistic Doppler boosting may also be consistent with the observed correlation between the peak frequency and isotropic luminosity of the flare \citep[e.g.][]{Buehler+2012}. Note that a relativistic Doppler boosting has been invoked in different models for the Crab flares, not necessarily based on turbulence \citep[e.g.][]{Yuan+2011, BednarekIdec2011, KomissarovLyutikov2011, Lyutikov+2012, ClausenLyutikov2012}.

In the turbulent scenario, the outer scale may be estimated from the duration of the flare, which gives
\begin{equation}
\label{eq:Lcrab}
L\sim ct_{\rm flare}\sim 2\times 10^{16}\left(\frac{t_{\rm flare}}{9{\rm\; days}}\right){\rm\; cm}\;.
\end{equation}
The inferred $L$ is comparable to the distance from the pulsar where the post-shock flow converges \citep[e.g.][]{Lyubarsky2012, Lyutikov+2018}. Since the total magnetic energy of the dissipation region, $\mathcal{E}\sim (4\pi L^3/3)\times (B^2/8\pi)$, is dissipated over a time scale $t_{\rm dyn}\sim L/c$, we can estimate the emitted isotropic power as $\mathcal{L}\sim \mathcal{E}/t_{\rm dyn}\sim cL^2B^2/6$. Using Eq. \eqref{eq:Lcrab} for $L$, we find
\begin{equation}
\label{eq:Bcrab}
B\sim 10^{-3}\left(\frac{\mathcal{L}}{4\times 10^{36}{\rm\; erg \; s}^{-1}}\right)^{1/2}\left(\frac{t_{\rm flare}}{9{\rm\; days}}\right)^{-1}{\rm\; G}\;.
\end{equation}
The required $B$ is consistent with the hypothesis that the flares are generated in the polar region. Indeed, in the polar region the magnetic field may be a factor of a few higher than the typical one within the bulk of the Nebula, where the magnetisation drops slightly below unity and $B\sim 1\mhyphen 3\times 10^{-4}{\rm\; G}$ \citep[e.g.][]{Hillas+1998, Meyer+2010}.

The typical Lorentz factor of the electrons emitting at the burn-off limit $\varepsilon_{\rm bo}$ is
\begin{equation}
\gamma_{\rm bo}\sim \left(\frac{e}{B\sigma_{\rm T}}\right)^{1/2}\sim 10^9\left(\frac{\mathcal{L}}{4\times 10^{36}{\rm\; erg \; s}^{-1}}\right)^{-1/4}\left(\frac{t_{\rm flare}}{9{\rm\; days}}\right)^{1/2}\;,
\end{equation}
where $\sigma_{\rm T}$ is the Thompson's cross section. The maximum Lorentz factor of the electrons is
\begin{equation}
\gamma_{\rm max}\sim \frac{eBL}{m_ec^2}\sim 10^{10}\left(\frac{\mathcal{L}}{4\times 10^{36}{\rm\; erg \; s}^{-1}}\right)^{1/2} \;.
\end{equation}
Since $\gamma_{\rm max}>\gamma_{\rm bo}$, in principle it is possible for the system to radiate photons at the burn-off limit.

Interestingly, for the Crab parameters $\gamma_{\rm bo}$ and $\gamma_{\rm max}$ are of the same order. This means that most of the flare energy is radiated by electrons moving at Lorentz factors close to the maximum allowed by the size of the system. Hence, the turbulent dissipation of the magnetic energy should heat most of the electrons up to $\gamma_{\rm max}$, which may happen when $\gamma_{\rm heat}\sim\gamma_{\rm max}$.\footnote{Since electrons cannot be accelerated up to energies much larger than $\gamma_{\rm heat}$ \citep[e.g.][]{Werner+2016, Kagan+2018}, models of the Crab flares based on reconnection of a Harris current sheet \citep[e.g.][]{Uzdensky+2011, Cerutti+2012, Cerutti+2013, Cerutti+2014} also require that $\gamma_{\rm heat}\sim\gamma_{\rm max}$. An interesting alternative possibility is explosive reconnection \citep[][]{Lyutikov+2017a, Lyutikov+2017b}. In this case, acceleration may produce two separate sub-populations of electrons, lowering the required magnetisation.} This condition can be expressed more transparently in terms of the pair multiplicity, $\kappa_e$. Suppose to have a relativistic MHD flow, with electromagnetic fields $E\sim B$ varying on a scale $L$. This requires the charge and current densities to be $\rho\sim j/c\sim B/L$. The number density of the free charges is $n_e\sim \kappa_e\rho/e\sim\kappa_e B/eL$, which using Eq. \eqref{eq:gamma_heat} gives
\begin{equation}
\gamma_{\rm heat}\sim\gamma_{\rm max}/\kappa_e\;.
\end{equation}
Hence, turbulence may power the Crab flares if the multiplicity in the dissipation region is extremely low, $\kappa_e\sim 1$ (for a similar suggestion on the multiplicity, see also \citealt{KirkGiacinti2017} and Section 4.5 of \citealt{Lyutikov+2018}).\footnote{The pair multiplicity may be indeed very low in the polar region of the Crab Nebula. As shown by recent PIC simulations of pulsar magnetospheres self-consistently including the process of pair creation, the pairs are created around the equatorial current sheet, while the solution may remain essentially charge separated at higher latitudes (e.g. \citealt{ChenBeloborodov2014, Philippov+2015b, Philippov+2015a, PhilippovSpitkovsky2018}; see \citealt{CeruttiBeloborodov2017} for a review).} When $\kappa_e\sim 1$, the system becomes starved of free charges immediately after turbulence starts to develop, and electrons may be accelerated up to $\gamma_{\rm heat}\sim\gamma_{\rm max}$.

Finally, note that the persistent spectrum of the Crab Nebula shows a possibly separate emission component in the gamma-rays, extending from photon energies of $\sim 1{\rm\;MeV}$ up to the burn-off limit \citep[e.g.][]{Dejager+1996, Vandermeulen+1998, Kuiper+2001}. \citet{Lyutikov+2019} suggested that this component is associated with the same emission site of the flares. An interesting possibility is that the entire emission above $\sim 1{\rm\;MeV}$ is due to the same electrons, which first produce the flares, and then emit at lower energies while cooling down.

\section{Conclusions}
\label{sec:conclusions}

We have presented a scenario for the dissipation of the magnetic energy in relativistic plasmas with a fast radiative (synchrotron and/or Inverse Compton) cooling. We have focused on plasmas where the particle number density is dominated by the pairs (electron-positron and electron-positron-ion plasmas). Inspired by the results of recent PIC simulations of particle heating in relativistic turbulence, we have suggested that:
\begin{itemize}
\item Electrons and positrons from the thermal background are impulsively heated up to $\gamma_{\rm heat}\sim \sigma_{0e}\sim B^2/8\pi n_em_ec^2$ by the strong electric field aligned with the local magnetic field that arises in intermittently formed current sheets. As an effect of different plasma compositions, $\gamma_{\rm heat}$ may change by a factor of a few.
\item As an effect of fast radiative cooling and of different injection times, an electron energy distribution of the form ${\rm d}n_e/{\rm d}\gamma\propto\gamma^{-2}$ is produced at $\gamma\lesssim\gamma_{\rm heat}$, while the distribution steepens and becomes dependent on the detailed properties of the plasma (e.g. magnetisation and amplitude of the magnetic field fluctuations compared to the background) at $\gamma\gtrsim\gamma_{\rm heat}$.
\item Since in the relativistic regime the Alfv\'en velocity is close to the speed of light, turbulence decays on the same time scale that photons take to escape. Hence, the magnetic and the radiation energy densities in the emission site are comparable.
\item Since most of the injected pairs are heated by a parallel electric field, their pitch angle is initially small. If the plasma has a significant proton component ($n_pm_p\gtrsim n_em_e$), pairs with a Lorentz factor smaller than $\gamma_{\rm iso}\sim m_p/m_e$ may be isotropised via gyro-resonant scattering \citep[][]{SobacchiLyubarsky2019}.
\end{itemize}
This scenario is very different with respect to the one proposed by \citet{Uzdensky2018}, where the electrons are heated slowly (instead of impulsively) and a quasi-thermal electron energy distribution forms as an effect of fast cooling. However, note that a quasi-thermal distribution may hardly be consistent with observations that require extended power-law electron energy distributions.

We have discussed the implications of the outlined scenario for blazar jets. Modelling the observed Spectral Energy Distribution indicates that the non-thermal electrons are distributed according a power-law of the form ${\rm d}n_e/{\rm d}\gamma\propto\gamma^{-2}$ for $\gamma\lesssim\gamma_{\rm b}$. Using the break Lorentz factor $\gamma_{\rm b}$ inferred from the SED as a proxy for the magnetisation of the pairs suggests that $\sigma_{0e}\sim 10^2$ in FSRQ and $\sigma_{0e}\gtrsim 10^3$ in BL Lacs. For a typical $n_e/n_p\sim 10\mhyphen 30$, the total (i.e. including the proton component) magnetisation of the plasma in the emission site is $\sigma_0\sim 1$ in FSRQ and $\sigma_0\gtrsim 15$ in BL Lacs. The effect of the anisotropy of the electron distribution might be important to model the SED of BL Lacs \citep[][]{SobacchiLyubarsky2019, TavecchioSobacchi2019}. The expected equipartition between the magnetic and the radiation energy density in the emission site may naturally explain why in BL Lacs the synchrotron and the Inverse Compton components of the SED have comparable luminosities.

Finally, we have examined the possibility that the gamma-ray flares from the Crab Nebula are powered by relativistic turbulence characterised by a $\sim$few light days length scale and a $\sim$mG magnetic field. During the flares most of the energy is radiated by pairs with a Lorentz factor close to the maximum allowed by the size of the system. We have shown that this requires the polar region of the Nebula, where the flares may be produced, to be characterised by a high degree of charge separation (equivalently, the pair multiplicity should be of order unity).

\section*{Acknowledgements}

We are grateful to Luca Comisso and Lorenzo Sironi for insightful discussions.
ES and YEL acknowledge support from the Israeli Science Foundation (grant 719/14) and from the German Israeli Foundation for Scientific Research and Development (grant I-1362-303.7/2016). ES also acknowledges support from NASA ATP NNX-17AG21G.

\def\aap{A\&A}\def\aj{AJ}\def\apj{ApJ}\def\apjl{ApJ}\def\mnras{MNRAS}\def\prl{Physical Review Letters}
\def\araa{ARA\&A}\def\physrep{PhR}\def\sovast{Sov. Astron.}\def\nar{NewAR}
\def\aapr{Astronomy \& Astrophysics Review}\def\apjs{ApJS}\def\nat{Nature}\def\na{New Astron.}
\def\prd{Phys. Rev. D}\def\pre{Phys. Rev. E}\def\pasp{PASP}\def\ssr{Space Sci. Rev.}
\bibliographystyle{mn2e}
\bibliography{2d}

\appendix
\section{Reconnection in charge starved current sheets}
\label{sec:appA}

Throughout the paper we have assumed that most of the turbulent energy is dissipated through reconnection of the anti-parallel component of the magnetic field across large-scale current sheets. Thin current sheets spontaneously form in MHD \citep[e.g.][]{Biskamp2003}: even if the thickness $\lambda$ of the current sheet is initially large with respect to the dissipation scale, the plasma inside the sheet is squeezed out with a velocity comparable to the Alfv\'en speed, and $\lambda$ may therefore reach the dissipation scale within a dynamical time.

The dissipation scale can be estimated as follows. From the Amp\`ere law, $\pa{\bf E}/\pa{\rm t}=c\nabla\times{\bf B}-4\pi{\bf j}$, one can estimate the current density inside the sheet as $j\sim cB/\lambda$, which increases as the current sheet is squeezing. When
\begin{equation}
\lambda\lesssim\lambda_{\rm starv}\equiv \frac{B}{e n_e}\;,
\end{equation}
the required current density exceeds $en_ec$, which is the maximum sustainable by the free charges that are present in the plasma. Since $\nabla\times{\bf B}$ is aligned with ${\bf B}$ in a force-free plasma, a parallel electric field starts growing once the current sheet becomes starved of free charges. Note that $\lambda_{\rm starv}\sim\sqrt{\sigma_{0e}}d_e\sim\sigma_{0e}\rho_e$, where $d_e\equiv\sqrt{m_ec^2/4\pi e^2 n_e}$ is the electron skin depth and $\rho_e\equiv m_e c^2/eB$ is the electron Larmor radius. In the relativistic regime $\sigma_{0e}\gg 1$, one finds that $\lambda_{\rm starv}\gg d_e\gg\rho_e$. Since the relativistic collisionless tearing instability grows on the time scale $t_{\rm tear}\sim(\lambda/d_e)^2 \times (\lambda/c)$ \citep[][]{ZelenyiKrasnoselskikh1979}, the tearing modes are still growing relatively slowly at the scale $\lambda\sim\lambda_{\rm starv}$, namely $t_{\rm tear}\sim \sigma_{0e}\times (\lambda_{\rm starv}/c)\gg \lambda_{\rm starv}/c$. Hence, we argue that the anti-parallel component of the magnetic field across the sheet reconnects when the current sheet becomes starved of free charges, which happens at the scale $\lambda\sim\lambda_{\rm starv}$.

\section{On the extension of the electron energy distribution}
\label{sec:appB}

As discussed in Section \ref{sec:spectrum}, in the fast cooling regime an electron energy distribution ${\rm d}n_e/{\rm d}\gamma\propto\gamma^{-2}$ is expected for $\gamma_{\rm cool}\lesssim\gamma\lesssim\gamma_{\rm heat}$. The acceleration of electrons at $\gamma\gtrsim\gamma_{\rm heat}$ depends on the ratio between the cooling time scale, $t_{\rm cool}\sim (\gamma_{\rm cool}/\gamma)\; t_{\rm dyn}$, and the acceleration time scale due to the stochastic scattering by the turbulent fluctuations, $t_{\rm acc}$. Using PIC simulations \citet{ComissoSironi2019} \citep[see also][]{Wong+2019} estimated that $t_{\rm acc}\sim t_{\rm dyn}/\sigma_e$, from which we find that
\begin{equation}
\frac{t_{\rm acc}}{t_{\rm cool}}\sim \frac{\gamma}{\gamma_{\rm cool}\sigma_e}\sim \frac{\gamma}{\gamma_{\rm cool}\gamma_{\rm heat}}\;.
\end{equation}
Acceleration by stochastic scattering becomes inefficient once $t_{\rm acc}/t_{\rm cool}\gtrsim 1$, namely when $\gamma\gtrsim\gamma_{\rm cool}\gamma_{\rm heat}$. Hence, in the presence of very efficient cooling (very low $\gamma_{\rm cool}$), the non-thermal electron distribution cannot extend up to Lorentz factors much larger than $\gamma_{\rm heat}$ \citep[for a similar conclusion see also][]{Zhdankin+2019}.

\end{document}